\documentclass[10pt]{article}
\usepackage{fullpage,citesort,epsfig,graphics,amsbsy}
\usepackage{psfrag}
\usepackage[normal,small]{caption}
\newcommand{\beq}{\begin{equation}}
\newcommand{\eeq}{\end{equation}}
\newcommand{\bea}{\begin{eqnarray}}
\newcommand{\eea}{\end{eqnarray}}

\newcommand{\be}{\begin{equation}}
\newcommand{\ee}{\end{equation}}
\newcommand{\bq}{\begin{eqnarray}}
\newcommand{\eq}{\end{eqnarray}}

\def\math{\mathsurround=0pt }
\def\leftrightarrowfill{$\math \mathord\leftarrow \mkern-6mu \cleaders\hbox{$\mkern-2mu \mathord- \mkern-2mu$}\hfill
 \mkern-6mu \mathord\rightarrow$}
\def\overleftrightarrow#1{\vbox{\ialign{##\crcr
     \leftrightarrowfill\crcr\noalign{\kern-1pt\nointerlineskip}
     $\hfil\displaystyle{#1}\hfil$\crcr}}}

\newcommand{\bfs}{\boldsymbol}

\let\l=\lambda

 \def\bd{\begin{document}} \def\ed{\end{document}}
\def\ds{\documentstyle} \let\fr=\frac \let\bl=\bigl \let\br=\bigr
\let\Br=\Bigr \let\Bl=\Bigl
\let\bm=\bibitem
\let\na=\nabla
\let\pa=\partial \let\ov=\overline
\def\ft#1#2{{\textstyle{{\scriptstyle #1}\over {\scriptstyle #2}}}}
\def\fft#1#2{{#1 \over #2}}
\def\vp{\varphi}
\def\sst#1{{\scriptscriptstyle #1}}
\def\oneone{\rlap 1\mkern4mu{\rm l}}
\def\td{\tilde}
\def\wtd{\widetilde}
\def\dalemb#1#2{{\vbox{\hrule height .#2pt
        \hbox{\vrule width.#2pt height#1pt \kern#1pt
                \vrule width.#2pt}
        \hrule height.#2pt}}}
\def\square{\mathord{\dalemb{6.8}{7}\hbox{\hskip1pt}}}
\def\wtd{\widetilde}
\def\R{\rlap{\rm I}\mkern3mu{\rm R}}
\def\im{{\rm i}}
\def\tilg{\tilde{g}}
\def\tilF{\tilde{F}}
\def\tilA{\tilde{A}}
\def\varf{\varphi}
\def\tilf{\tilde{\phi}}
\def\tilh{\tilde{h}}
\def\rme{{\rm e}}
\def\ep{\epsilon}
\def\0{{(0)}}
\def\9{{(9)}}
\def\8{{(8)}}
\def\7{{(7)}}
\def\6{{(6)}}
\def\5{{(5)}}
\def\4{{(4)}}
\def\3{{(3)}}
\def\2{{(2)}}
\def\1{{(1)}}
\newcommand{\trace}{{\rm Tr}}
\newcommand{\ub}{\overline{U}}
\newcommand{\vb}{\overline{V}}
\newcommand{\uh}{\widehat{U}}
\newcommand{\vh}{\widehat{V}}
\newcommand{\ubh}{\overline{\widehat{U}}}
\newcommand{\vbh}{\overline{\widehat{V}}}
\newcommand{\lb}{\bar{\l}}
\newcommand{\Fb}{\overline{F}}
\newcommand{\Fh}{\widehat{F}}
\newcommand{\Fbh}{\overline{\widehat{F}}}
\newcommand{\Ab}{\overline{A}}
\newcommand{\Ah}{\widehat{A}}
\newcommand{\Abh}{\overline{\widehat{A}}}
\newcommand{\Gb}{\overline{G}}
\newcommand{\Gh}{\widehat{G}}
\newcommand{\Gbh}{\overline{\widehat{G}}}
\newcommand{\Pb}{\overline{P}}
\newcommand{\Ph}{\widehat{P}}
\newcommand{\Pbh}{\overline{\widehat{P}}}
\newcommand{\Qb}{\overline{Q}}
\newcommand{\Qh}{\widehat{Q}}
\newcommand{\Qbh}{\overline{\widehat{Q}}}
\newcommand{\Bb}{\overline{B}}
\newcommand{\Bh}{\widehat{B}}
\newcommand{\Bbh}{\overline{\widehat{B}}}
\newcommand{\fhns}{\hat{F}^{\rm (NS)}}
\newcommand{\fhrr}{\hat{F}^{\rm (RR)}}
\newcommand{\ahns}{\hat{A}^{\rm (NS)}}
\newcommand{\ahrr}{\hat{A}^{\rm (RR)}}
\newcommand{\hhrr}{\hat{H}^{\rm (RR)}}
\newcommand{\hchi}{\hat{\chi}}
\newcommand{\hphi}{\hat{\phi}}
\newcommand{\htau}{\hat{\tau}}
\newcommand{\cG}{{\cal G}}
\newcommand{\cGb}{\overline{{\cal G}}}
\newcommand{\cH}{{\cal H}}
\newcommand{\cP}{{\cal P}}
\newcommand{\cPb}{\overline{{\cal P}}}
\newcommand{\cQ}{{\cal Q}}
\newcommand{\cQb}{\overline{{\cal Q}}}
\newcommand{\cM}{{\cal M}}
\newcommand{\cN}{{\cal N}}
\newcommand{\cO}{{\cal O}}
\newcommand{\cD}{{\cal D}}
\newcommand{\cL}{{\cal L}}
\newcommand{\cA}{{\cal A}}
\newcommand{\cB}{{\cal B}}
\newcommand{\hg}{\hat{g}}
\newcommand{\cE}{{\cal E}}

\newcommand{\vpp}{\mbox{$\langle{\scriptstyle++}\rangle$}}
\newcommand{\vmp}{\mbox{$\langle{\scriptstyle-+}\rangle$}}
\newcommand{\vppp}{\mbox{$\langle{\scriptstyle+++}\rangle$}}
\newcommand{\vmpp}{\mbox{$\langle{\scriptstyle-++}\rangle$}}
\newcommand{\vpmp}{\mbox{$\langle{\scriptstyle+-+}\rangle$}}

\begin{document}
\setlength{\captionmargin}{20pt}
\begin{flushright}
UFIFT-HEP-03-24\\
hep-th/0310121
\end{flushright}

\vskip0.3cm

\begin{center}
\begin{Large}
{\bf Quantum Field Theory in the Language of Light-cone 
String\footnote{Supported 
in part by the Department
of Energy under Grant No. DE-FG02-97ER-41029. 
}}
\end{Large}

\vskip0.3cm
{\large 
 Charles B. Thorn\footnote{E-mail  address: {\tt thorn@phys.ufl.edu}}
}
\vskip0.30cm
{\it Institute for Fundamental Theory, Department of Physics\\ 
University of Florida, Gainesville FL 32611}


\vskip0.2cm
\end{center}

\begin{abstract}\noindent
The worldsheet representation of the sum of the planar diagrams
of scalar $\Phi^3$ field theory and ${\cal N}=0,1,2,4$
supersymmetric Yang-Mills theory is explained. This
was a talk given to the Light Cone Workshop: Hadrons
and Beyond, 5-9 August 2003, University of Durham.
\end{abstract}
%

\section{Introduction}
One of the most striking dualities that has emerged in the
development of string theory is the so-called AdS/CFT correspondence
\cite{klebanov,maldacena} in which the $N_c=\infty$ limit of
the conformal invariant
${\cal N}=4$ extended supersymmetric Yang-Mills theory is
conjectured to be equivalent to a noninteracting IIB
superstring theory on an AdS$_5\times$S$_5$ background.
Although this duality is not yet proven, it is
supported by an impressive amount of evidence. 
If true it is a stunning breakthrough in our thinking
about string theory: it strongly supports 
the idea that all of the apparently non-local features of
string (or string field theory) are simply due to an
awkward choice of variables. In other words, there should be an
alternative manifestly local formulation (in this case the SUSY 
Yang-Mills theory) underlying string dynamics \cite{thornmosc}. 

On the other hand the duality also points to a breakthrough in
our thinking about quantum field theory. The possibility
of a stringy formulation of field theory could provide a powerful
new way to understand hadron spectroscopy including
quark confinement \cite{klebanovs,polchinskis}. 
After all, one of the most compelling
mechanisms for quark confinement is the formation of
color flux confining tubes between separated color sources, and
a stringy reformulation of QCD might be just the long-sought
change of variables that definitively clarifies the origin of
these flux tubes.
 
Following this line of thought, Bardakci and I \cite{bardakcit}
were motivated to build a worldsheet representation of the
sum of planar diagrams in a generic quantum field theory.
In contrast to the AdS/CFT correspondence, for which the
the string side is well understood only when the field theory 
coupling is large, our construction bases its string description
directly on the weak coupling expansion of the field theory. 
It thus shows that a worldsheet interpretation of field
theory is generic and by no means limited to the very
special circumstances of the AdS/CFT duality.

In our method we first find a worldsheet
description of each planar Feynman diagram parameterized with
light-cone variables. Then the stringy description we build 
comes out in terms of the light-cone worldsheet familiar in
string theory \cite{goddardgrt}. Once each diagram is 
given a worldsheet description, the sum of all planar
diagrams can be formulated directly on that worldsheet
template, producing a string theory in which the target space
variables ${\boldsymbol q}(\sigma,\tau)$ interact with an
Ising spin system living on the worldsheet template. It
seems that these Ising spins play a role analogous
to that of the AdS fifth dimension.

\section{Lightcone String Basics}
Let us begin by recalling the description of the lightcone worldsheet
in the bosonic string theory \cite{goddardgrt}.
It is most convenient to use a phase space action principle
with coordinates and conjugate momenta
$x^\mu(\sigma,\tau), {\cal P}^\mu(\sigma,\tau)$. Then
the lightcone parameterization of the worldsheet is easily specified:
$x^+=(x^0+x^3)/\sqrt2=\tau$, 
and ${\cal P}^+=({\cal P}^0+{\cal P}^3)/\sqrt2=1$. 
Since ${\cal P}^\mu$ is the density of
energy momentum on the string, the range of $\sigma$ is 
$0<\sigma<p^+$ where $p^+$ is the total $+$ component of momentum.
Then the action for the dynamics of string is simply 
\bea
S&=&\int d\tau \int_0^{p^+} d\sigma \left(\dot{\boldsymbol x}\cdot{\boldsymbol 
{\cal P}}-{1\over2}{\boldsymbol{\cal P}}^2
-{T_0^2\over2}{\boldsymbol x}^{\prime2}\right)
\eea
One gets to the more familiar configuration space action by algebraically
eliminating ${\boldsymbol{\cal P}}$:  
\bea
S&\to&\int d\tau \int_0^{p^+} d\sigma 
{1\over2}\left({\dot{\boldsymbol{x}}}^2-T_0^2{\boldsymbol x}^{\prime2}
\right).
\eea
For the path history version of the dynamics of string it is convenient
to work with a Euclidean worldsheet, which means that we continue to
imaginary $\tau$, $i\tau\to\tau>0$. Then the exponent in
the path integrand becomes
\bea
iS&\to&-\int d\tau \int_0^{p^+} d\sigma 
{1\over2}\left({\dot{\boldsymbol{x}}}^2+T_0^2{\boldsymbol x}^{\prime2}
\right).
\eea
Finally, it will turn out that our worldsheet description of
the Feynman diagrams of field theory will require the use of
dual target space variables ${\boldsymbol q}(\sigma,\tau)$ 
related to the transverse coordinates by
$({\boldsymbol q}^\prime,{\dot{\boldsymbol q}})
=({\dot{\boldsymbol x}},-{\boldsymbol x}^\prime/T_0^2)$. The
open string boundary conditions ${\boldsymbol x}^\prime=0$ 
then go to Dirichlet conditions ${\dot{\boldsymbol q}}=0$,
with boundary values satisfying 
${\boldsymbol q}(p^+,\tau)-{\boldsymbol q}(0,\tau)={\boldsymbol p}$,
the total transverse momentum carried by the string.
In these variables the Euclidean worldsheet action becomes
\bea
iS&=&-\int d\tau \int_0^{p^+} d\sigma 
{1\over2}\left({\boldsymbol{q}}^{\prime2}+T_0^{-2}{\dot{\boldsymbol q}}^{2}
\right).
\eea
We shall find a similar representation for the free field theory
propagator except that the time derivative term will be absent. This
is reasonable since the point particle limit of string is $T_0\to\infty$.

\section{QFT Lightcone Worldsheet}
Now we turn to the worldsheet representation of the individual
planar diagrams of a field theory with cubic couplings. We
start with the free propagator. 
In the mixed $(\tau=ix^+,p^+,{\boldsymbol p})$ representation,
the Feynman propagator for a massless scalar with $p^+>0$ 
is just $\theta(\tau)e^{-\tau{\bfs p}^2/2p^+}/2p^+$.  
We choose to absorb the $1/2p^+$ factor in the 
{\it earlier} vertex to which the propagator attaches. 
Then in \cite{bardakcit}, we
note the following remarkable identity
\begin{eqnarray}
\exp\left\{-{T\over2p^+}
{{\boldsymbol{p}}^2}\right\}&=&
\int_{{{\boldsymbol{q}}(0)=0\atop
{{\boldsymbol{q}}(p^+)={\boldsymbol{p}}}}} 
DcDbD{\boldsymbol{q}}\ e^{-S_0}\\
S_0&=&\int_0^T d\tau\int_0^{p^+} d\sigma
\left({1\over2}
{\boldsymbol q}^{\prime2}-
{\bfs b}^\prime {\bfs c}^\prime\right)
\end{eqnarray}
where it is understood that $\boldsymbol{b}=\boldsymbol{c}=0$
on all boundaries.
This formula is central to our construction. The right side is 
a path integral over bosonic target space variables ${\bfs q}(\sigma,\tau)$
together with Grassmann variables 
${\bfs b}(\sigma,\tau), {\bfs c}(\sigma,\tau)$
defined on a rectangular worldsheet of dimensions $p^+\times T$
just like the worldsheet of the lightcone string propagator.
If $D=d+2$ is the spacetime dimension then ${\bfs q}$ has
$d$ components and ${\bfs b}$, ${\bfs c}$ each have $d/2$
components. The purpose of the Grassmann variables is
to cancel the determinant factors arising from the ${\bfs q}$
integration.
The $p^+$ in the denominator of the exponent on the
left side appears only in the geometrical width of the
worldsheet on the right side. The ${\bfs p}$ dependence on the
left side appears only in the Dirichlet boundary conditions
on the right side.
In effect the formula represents a 
field quantum as a composite of string bits
if we discretize  $p^+=Mm=$(Number of bits)$\times m$. 
Each bit carries a single unit $m$ of $p^+$. 

To give rigorous meaning to the
path integral on the right side it is natural \cite{gilest} to put
$(\sigma,\tau)$ on a rectangular lattice of size $M\times N$.
Here $M=p^+/m$ is the longitudinal momentum in units of $m$
and $N=T/a$ is the evolution time in units of $a$. This worldsheet
grid is the template on which the sum over all planar diagrams
is to be performed. On each site there will be target
space variables ${\bfs q}_i^j,{\bfs b}_i^j,{\bfs c}_i^j$,
and the functional integrals over them are just ordinary
integrals on the lattice.

Diagrams with $n$ loops are represented by $n$ line segments 
extended in time but
at fixed $\sigma$, just as in
Mandelstam's interacting string diagrams \cite{mandelstam}. 
The location and length of each of these segments
is summed. These line segments represent internal boundaries on which
Dirichlet conditions are imposed with a different value for
${\bfs q}$ on each segment. The ${\bfs q}$ on
each internal segment is independently integrated.
As an example we draw the one loop self energy diagram below.
The internal boundary representing the loop 
is indicated by the solid line segment. The dotted lines
represent the {\it absence} of a boundary.
\begin{center}
\psfrag{'k1'}{$k_1=k_0+k$}
\psfrag{'k0'}{$k_0$}
\psfrag{'l'}{$l$}
\psfrag{'M'}{$M$}
\includegraphics[width=6cm]{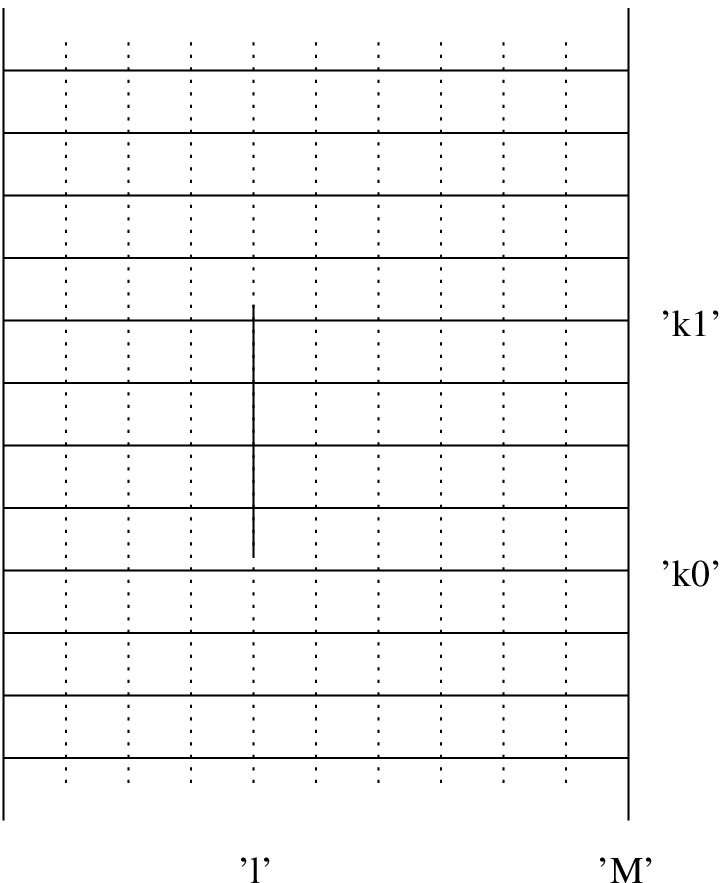}
\end{center}
\noindent The loop in this figure has length $ka$ and its earlier
end is located at $\sigma=lm$ and $\tau=k_0$.
The boundary value of ${\bfs q}$ on the
internal solid line is integrated over all real values.

Consider next a general multi-loop diagram for a field theory with
cubic couplings:
\begin{center}
\includegraphics[width=5.5cm]{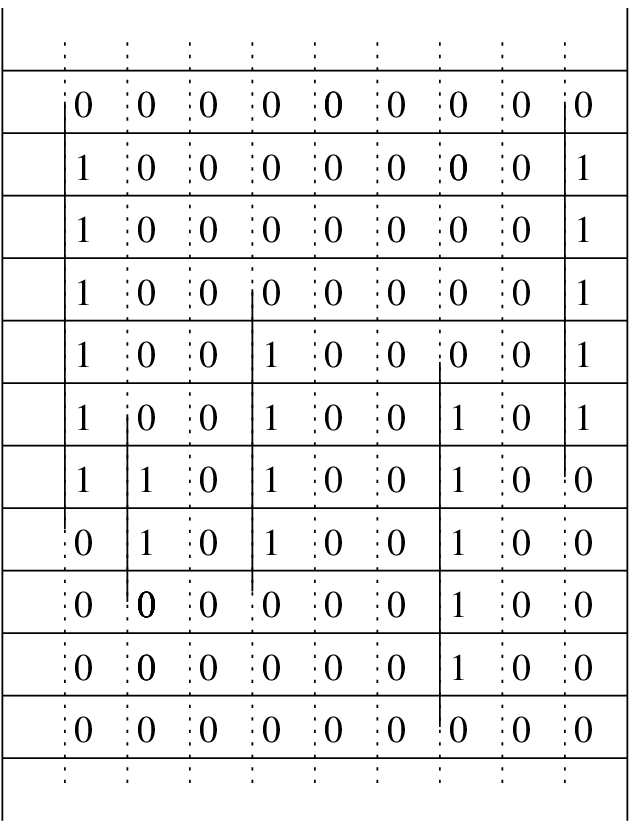}
\end{center}
\noindent We observe that the structure of the diagram is completely
characterized by telling for each site whether or not it
is crossed by a boundary. We can keep track of this information
by introducing a two valued Ising spin $s_i^j$ for each site 
with value $+1$ if the site is crossed by a boundary and
with value $-1$ if it is not. It is also convenient
to use the spin up projector $P_i^j=(1+s_i^j)/2$ with corresponding
values $1,0$. The values $1,0$ indicated on each temporal
link $(i,j)(i,j+1)$ of the above diagram are just the values of 
$P_i^jP_i^{j+1}$. The summation over all planar diagrams is just
the sum over all spin configurations appropriately weighted.

The appropriate weight to use in the path integral can
be read off from the usual Feynman rules. It is immediate
that in the ``bulk'', away from boundaries, the appropriate
weight is just $e^{-S_0}$. On the boundaries the weighting
has to enforce Dirichlet boundary conditions. For each
pair of consecutive sites crossed by a boundary there must
be a delta function that forces the ${\bfs q}$'s on the
two sites to be equal. The same effect can be obtained by
adding a term $\sum_{i,j}({P_i^jP_i^{j-1}a/2m\epsilon})({\boldsymbol q}_i^j
-{\boldsymbol q}_i^{j-1})^2$ to the action with the understanding
that $\epsilon\to0$ eventually. Note that such a term brings
into the action an effective $\dot{\bfs q}^2$ whose
coefficient ${P_i^jP_i^{j-1}a^2/m^2\epsilon}\sim {T_{eff}^{-2}}(i,j)$,
where $T_{eff}$ is an effective locally dynamical string tension.
This is quite like the coupling of the target space to the
AdS radial coordinate in the AdS/CFT correspondence.  

\section{Worldsheet System for $\Phi^3$ Field Theory}
The details of the worldsheet system that sums the
planar diagrams of a field theory depend on the theory.
We give here the one that sums the {\it bare} planar
diagrams of a massive scalar field theory with only
cubic interactions $g\Phi^3$.
\begin{eqnarray}
T_{fi}&=&\lim_{\epsilon\to0}
\sum_{s_i^j=\pm1}\int DcDbD{\boldsymbol q}
\exp\left\{\ln{\hat g}\sum_{ij}{1-s_i^js_i^{j-1}\over2}
-{d\over2}\ln\left(1+\rho\right)\sum_{i,j}P_i^j\right\}
 \nonumber\\
&&\hskip-.3in\exp\left\{-{a\over2m}\sum_{i,j}
{({\boldsymbol q}_{i+1}^j-{\boldsymbol q}_{i}^{j})^2}
-{a\over2m\epsilon}\sum_{i,j}P_i^jP_i^{j-1}
{({\boldsymbol q}_{i}^j-{\boldsymbol q}_{i}^{j-1})^2}
\right\}\\ 
&&\hskip-.3in\exp\left\{{a\over m}
\sum_{i,j}\left[A_{ij}{\boldsymbol b}^j_{i}{\boldsymbol c}^j_{i}
+C_{ij}({\boldsymbol b}_{i+1}^j-{\boldsymbol b}_i^j)
({\boldsymbol c}_{i+1}^j-{\boldsymbol c}_i^j)
-B_{ij}{b}_{i}^{j}{c}_{i}^{j}
-D_{ij}({b}_{i+1}^j-{b}_i^j)({c}_{i+1}^j-{c}_i^j)\right]\right\}
\nonumber
\eea
On the lattice the functional measure in this formula is 
rigorously defined as the
product of many ordinary integration measures
\bea
DcDbD{\boldsymbol q}&\equiv&
\prod_{j=1}^N\prod_{i=1}^{M-1} 
{d{\boldsymbol c}^j_id{\boldsymbol b}^j_i\over2\pi} 
d{\boldsymbol q}^j_i
\eea
We have also employed both forms of the Ising spin variables
$s_i^j$ and $P_i^j=(1+s_i^j)/2$. We have used the ratio of
lattice constants $m/a$ to define a dimensionless coupling
constant ${\hat g}>0$ by ${\hat g}^2
= (g^2/64\pi^3) \left({m/2\pi a}\right)^{(d-4)/2}$. According to
our worldsheet picture there should be a factor of ${\hat g}$
at the beginning and end of each internal solid line. In terms of the
Ising spin configuration, these points are where a spin flips.
Thus the first term in the first exponent provides exactly the
right factors of coupling for every planar diagram. The first
term in the second exponent is just the ${\bfs q}$ part of
$S_0$. The second term in the second exponent enforces
Dirichlet boundary conditions on the ${\bfs q}$ variables
as already discussed.

The terms in the third exponent require some explanation. 
These terms all involve the Grassmann ghosts. Although
there are many terms that we will describe in a moment,
note that the Grassmann integrations on different time slices are decoupled
from one another.
The coefficients $(A, B, C, D)_{i,j}$ are each polynomials in the
$P$'s associated with lattice sites at most two steps away from
site $(i,j)$. Thus the entire expression defines a worldsheet
system with completely local dynamics. These
coefficients are defined in detail as follows:
\bea
&&\hskip-.4in A_{ij}={1\over\epsilon}{P}_i^j{P}_i^{j-1}
+{P}_i^{j+1}{P}_i^j-{P}_i^{j-1}{P}_i^j{P}_i^{j+1}+
(1-P_i^j)(P_{i+1}^j+P_{i-1}^j)+\rho(1-P_i^j)P_{i-1}^{j-1}P_{i-1}^j\\
&&\hskip-.4in
B_{ij}=(1-P_i^j)P_i^{j-1}P_i^{j-2}P_{i+1}^j+
(1-P_i^j)\left(P_{i+1}^jP_{i+1}^{j+1}
(1-P_{i+1}^{j-1})+
P_{i-1}^j{P_{i-1}^{j+1}(1-P_{i-1}^{j-1})}\right)\\
&&\hskip-.4in C_{ij}=(1-P_i^j)(1-P_{i+1}^j)\\
&&\hskip-.4in D_{ij}=(1-P_i^j)(1-P_{i+1}^j)P_i^{j-1}P_i^{j-2}
\eea
The $1/\epsilon$ term in $A$ is exactly correlated with
the second term in the second exponent and together they provide
the properly normalized delta function in the limit $\epsilon\to0$
that enforces Dirichlet conditions on the ${\bfs q}$'s.
The parameter $\rho=\mu^2a/(md-\mu^2a)$ appearing in the
last term of the first exponent and the last term in $A$
gives a mass $\mu$ to the scalar field. The $C$ terms are
precisely the ghost terms in $S_0$ that involve differences
of ghost fields at adjacent sites. The spin projectors 
in $C$ kill their contribution when one of
the adjacent sites has spin $+1$, i.e. when it is on a boundary. 
The remaining ghost terms in $S_0$ are supplied by the fourth
term in $A$. Without the second and third terms of $A$, the path integrand
would be independent of the Grassmann variables on the
earliest site on each internal solid line. 
Their presence is solely to  prevent the integrations over
those variables from giving zero! 

Although the expressions for $B$ and $D$ look rather intimidating,
inspection of them shows that they are designed to strategically
cancel certain terms in $A$ and $C$ respectively. Note that
they only depend on one of the components of ${\bfs b}$, ${\bfs c}$,
written in non-bold type. Their effect is to provide the
$1/p^+$ factors which we removed from the propagators and
absorbed in the earlier vertex. The remarkable observation
here is that these apparently nonlocal factors are supplied
by a {\it local} modification of the ghost action. Indeed that
is the profound point about the formula: the entire sum of
lightcone parametrized planar diagrams is produced by a local
world sheet dynamics, in spite of the prolific number of
rational functions of the $p^+$'s that infest the usual
representation of the diagrams. I hasten to stress that the
formula is strictly valid only for the bare diagrams and
is complete only in space-time dimensions less than $4$
for which ultraviolet divergences don't generate 
violations of Lorentz invariance. The full power of
string theory can only be unleashed on these field theories
after it is demonstrated that any counter-terms
required to restore Lorentz invariance 
also have a local worldsheet description.

\section{The Planar Yang-Mills Worldsheet}
Scalar $\Phi^3$ theory is fine as a laboratory for developing
the worldsheet formalism, but we are really interested in applying
the method to QCD. As 't Hooft pointed out long ago \cite{thooftlargen} the
planar diagram approximation to QCD is singled out by
the $N_c\to\infty$ limit, which also suppresses internal quark
loops. Thus one can focus on the worldsheet construction that
sums the planar diagrams of pure Yang-Mills theory. Actually
for glueball spectroscopy the large $N_c$ limit generalizes 
the dominant ``planar'' diagrams to all those that can be 
drawn on a cylinder with no crossed lines. Given our
lightcone methodology it is natural to work in lightcone gauge $A^+=-A_-=0$.
Then one eliminates $A_+$ by solving the constraints and
is left with Feynman rules for the transverse fields ${\bfs A}$
only. Here we restrict attention to four space-time dimensions
and then it is convenient to use a complex basis 
$A_\wedge=(A_1+iA_2)/\sqrt2$, $A_\vee=(A_1-iA_2)/\sqrt2$
which is depicted by attaching an arrow to the propagator line.
With this notation the only non-vanishing planar vertices are:
\bea
{{}\atop\mbox{\includegraphics[width=1.5cm]{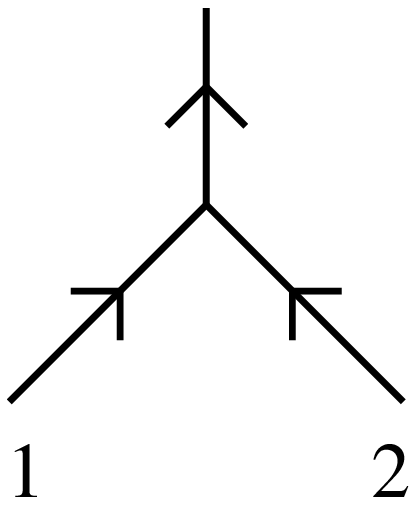}
}}
\displaystyle\quad&=&{{ga\over 4m\pi^{3/2}}p^+_3
\left({p_2^\wedge\over p^+_2}-{p_1^\wedge\over p^+_1}\right)}\\
{{}\atop\mbox{\includegraphics[width=1.5cm]{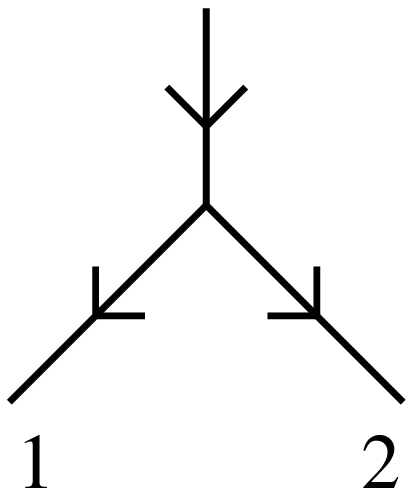}
}}
\displaystyle\quad&=&{{ga\over 4m\pi^{3/2}}p^+_3
\left({p_2^\vee\over p^+_2}-{p_1^\vee\over p^+_1}\right)}
\eea
where the momenta are understood to be flowing into the
vertex. The new features that must be brought into the
worldsheet construction of the previous section are
(1) new worldsheet degrees of freedom to account for the flow of
spin (polarization) through an arbitrary planar diagram, and
(2) a way to deal with the momentum dependence of the
vertex. The first can be handled by introducing Neveu-Schwarz
like fermionic variables, but I refer the reader to the original
paper \cite{thornsheet} for details. Here we concentrate on
the issue of momentum dependence.

We must produce the momentum factors by some local
feature of the worldsheet formalism. The key is to
consider the expectation value of ${\bfs q}^\prime(\sigma,\tau)$
at some point on the worldsheet of the free propagator.
Using its discretized form we find \cite{thornsheet} 
\begin{eqnarray}
{1\over m}\langle{\boldsymbol q}_l-{\boldsymbol q}_{l-1}\rangle
={{\boldsymbol q}(p^+)-{\boldsymbol q}(0)\over p^+}={{\bfs p}\over p^+}
\end{eqnarray}
which is exactly one of the terms we want to generate. Since
the expectation is independent of location as long as
it is an interior point on the propagator world sheet we are free to place
such an insertion in the neighborhood of the end of
the internal boundary that marks the spot where the gluon
fission or fusion occurs. We choose candidate locations
as marked by the open circles in the following worldsheet diagrams
for the cubic vertex:

\begin{center}
\psfrag{'A'}{$\circ$}
\psfrag{'B'}{$\circ$}
\psfrag{'C'}{$\circ$}
\psfrag{'D'}{}
\psfrag{'l'}{}
\psfrag{'k'}{}
\psfrag{'1'}{1}
\psfrag{'2'}{2}\psfrag{'3'}{3}
\includegraphics[width=10cm]{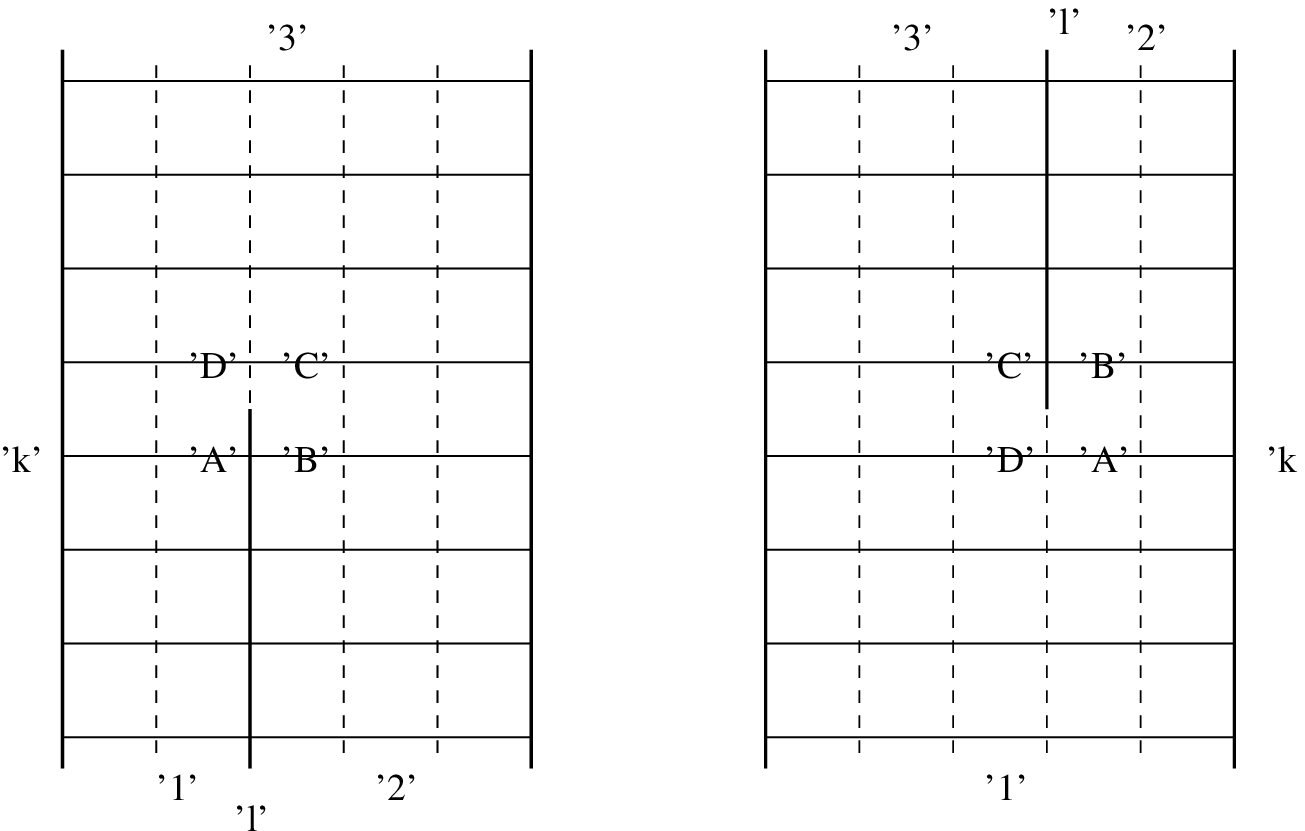}
\end{center}

\noindent We control which combination of factors is
produced by choosing the circle that lies on the appropriate 
gluon propagator coming into the vertex. This method 
locally produces everything in the vertex except the overall factor
of $p^+_3$. In \cite{thornsheet} I showed how another set of Grassmann
variables can be designed to locally produce these factors.

The final thing I have to say is that the worldsheet
formalism described above automatically includes the quartic Yang-Mills
vertex!
To see how, it is enough to consider a worldsheet with two cubic vertices.
Each vertex will have some combination of the ${\bfs q}^\prime$
insertions just discussed. When the two insertions
are at different times or on different gluon propagators
nothing changes: one just has the value of the diagram with two
cubic vertices. But when the two insertions are on the same
time slice and on the same propagator, there is of course a fluctuation 
contribution to the expectation:
\bea
{1\over m^2}\left\langle (q_{i+1}-q_i)(q_{j+1}-q_j)\right\rangle
=\left({q(p^+)-q(0)\over p^+}\right)^2
+{1\over a}\left[{1\over m}\delta_{ij}-{1\over p^+}\right]
\eea
The two situations in which this fluctuation term comes into play
are illustrated in the following diagrams. The open squares
show where the insertions must be to yield a fluctuation
contribution. On the left is the coincidence limit
of two cubics in a $t$ (exchange) channel diagram while on the
right the coincidence limit is in an $s$ (direct) channel diagram.
Remarkably the combination of the $-{1/ap^+}$ fluctuation terms from
the two contributions exactly reproduces the Yang-Mills quartic
vertex! We know that the quartic vertex is required by gauge invariance, 
but apparently the worldsheet formalism is clever
enough to know about this subtle requirement {\it and} to
achieve it {\it locally} on the worldsheet. I regard this as a
dramatic indication that our way of building a worldsheet
interpretation is definitely on the right track. 
The rest of the fluctuation term $\delta_{ij}/am$ term is 
not directly associated with the quartic vertex but in any case is
a local $\delta(\tau)\delta(\sigma)$ worldsheet contact term,
whose role is still mysterious.
\begin{center}
${\includegraphics[width=4.5cm]{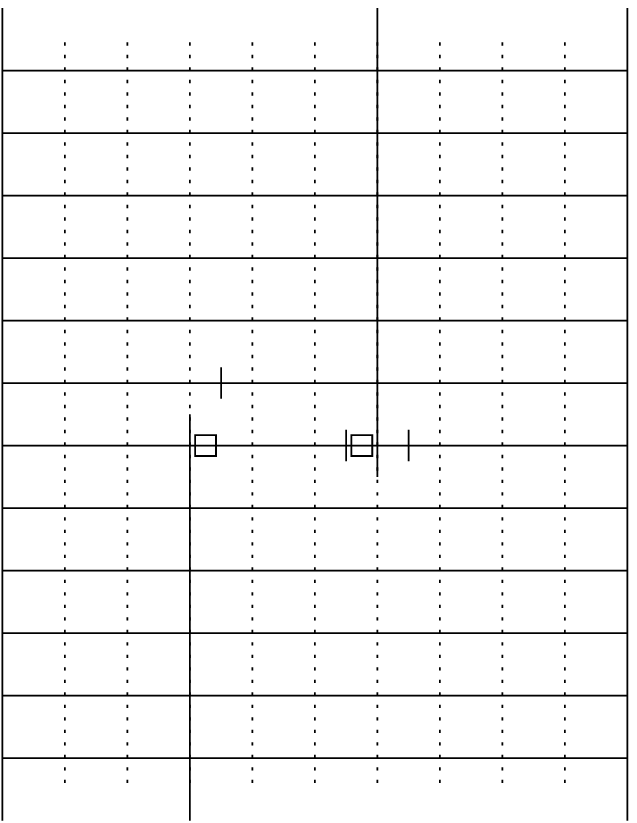}}
\hskip1cm\quad
{\includegraphics[width=4.5cm]{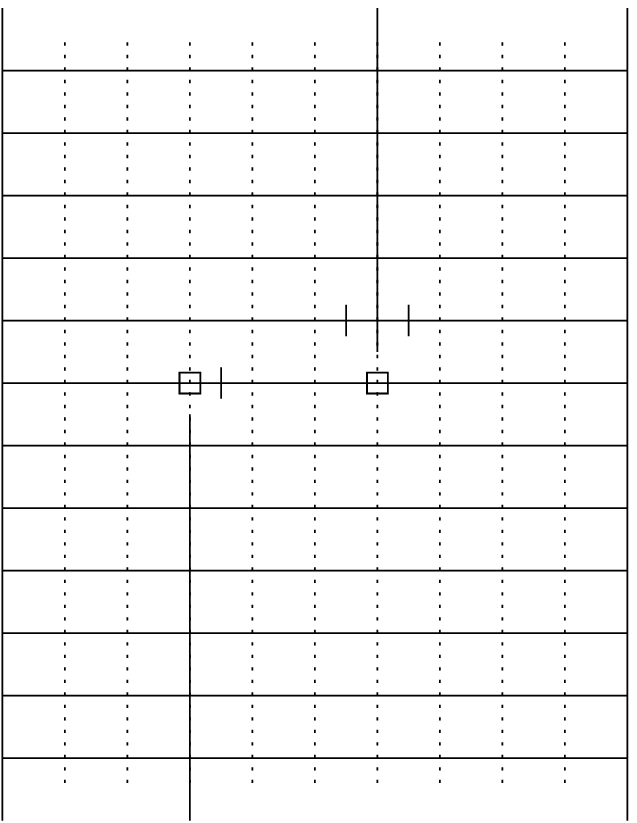}}
$
\end{center}

\section{Supersymmetry}
I remark briefly that the worldsheet construction has
been extended to ${\cal N}=1,2,4$ supersymmetric Yang-Mills theories
\cite{gudmundssontt}.
For ${\cal N}=1$ SUSY one adds the appropriate fermions,
and everything works as with pure Yang-Mills.
For ${\cal N}=2,4$ SUSY one first adds 2 or 6 ``dummy dimensions'',
freezing the values of the extra components of ${\bfs q}$ to be
zero on all worldsheet boundaries, internal as well as external. 
Then the extra ghost integrals exactly cancel the 
extra ${\bfs q}$ integrals. Then it turns out that the 
fluctuations of these dummy-dimensions generate all the
required quartic interactions of extended SUSY, {\it exactly} as in pure
Yang-Mills.
\section{Conclusion: Recent and Ongoing  Work}
I hope I have managed here to convey the basic
principles and methods underlying the worldsheet description
of quantum field theory developed by Bardakci and
me. We have constructed a worldsheet ``template'' for 
summing the planar diagrams in a broad range of interesting theories.
There has not been time to describe several
approaches we have taken toward using the formalism
as a calculational tool. In 
\cite{bardakcitmean,bardakcitimp,thorntfisheet,bardakci}
we have developed several versions of a mean field approximation to
the dynamics of the worldsheet Ising spin system. 
Results for scalar $\Phi^3$ theory 
indicate a regime at strong coupling where the
mean field for the Ising spin plays a role similar to that
of the AdS radial coordinate in the AdS/CFT correspondence.
A similar analysis for planar QCD and SUSY gauge theories 
remains to be done.

The important issue of renormalization in the worldsheet
formalism is wide open and under active investigation. 
We have emphasized that
the worldsheet systems we construct reproduce all {\it bare}
planar diagrams, which means it can be completely
trusted in space-time dimensions sufficiently low to ensure the absence of 
ultraviolet divergences in the field theory. 
The worldsheet lattice we use
actually cuts off all divergences, but not in a manifestly Lorentz
invariant away. Consequently field theoretic ultraviolet divergences
produce Lorentz violating artifacts that survive the
continuum limit. Counter-terms must therefore be introduced
to cancel these artifacts. Glazek has constructed a set
of counter-terms for lightcone quantum field theory \cite{glazek} but it
is not obvious from his work that counter-terms can be chosen as
{\it local} worldsheet modifications. Indeed, I think
it is fair to say that proving
that all necessary counter-terms for restoring
Lorentz invariance are local on the worldsheet is tantamount to
definitively establishing our worldsheet system as a bona
fide representation of the fully renormalized Yang-Mills theory
in four space-time dimensions. 
\section{Acknowledgments}
I am grateful to K. Bardakci, S. Glazek, J. Greensite,
S. Gudmundsson, and T. Tran for valuable discussions.
This work was supported by in part by the Department
of Energy under Grant No. DE-FG02-97ER-41029.

\end{document}